\documentclass[aip, amsmath, amssymb,reprint]{revtex4-1}    
\usepackage{bm}
\usepackage{graphicx}
\usepackage[utf8]{inputenc}
\usepackage[T1]{fontenc}
\usepackage{mathptmx}
\usepackage{hhline}

\newcommand{\dd}{\partial}
\newcommand\mycom[2]{\genfrac{}{}{0pt}{}{#1}{#2}}
\def \de{\delta}

\def \eps{\varepsilon}

\begin{document}

\title{Magnetic energy spectrum produced by turbulent dynamo: effect of time irreversibility.
}

\author{ A.V. Kopyev}
\email{kopyev@lpi.ru}
\author{ A.S. Il'yn}
\altaffiliation[Also at ]{National Research University Higher School of Economics, 101000,
Myasnitskaya 20, Moscow, Russia}
\email{asil72@mail.ru}
\author{V.A. Sirota}
\email{sirota@lpi.ru}
\author{K.P. Zybin}
\altaffiliation[Also at ]{National Research University Higher School of Economics, 101000,
Myasnitskaya 20, Moscow, Russia}
\email{zybin@lpi.ru}
\affiliation{P.N.Lebedev Physical Institute of RAS, 119991, Leninskij pr.53, Moscow,
Russia}


\begin{abstract}
We consider the kinematic stage of evolution of magnetic field advected by
turbulent hydrodynamic flow. We use  a generalization of the Kazantsev-Kraichnan model to
investigate time irreversible flows. In the viscous range of scales, the infinite-time limit of the
spectrum is a power law but its slope is more flat than that predicted by Kazantsev model. 
This result agrees  with  numerical simulations. 
The rate of magnetic energy growth is slower than that in the time-symmetric case. We show that for
high magnetic Prandtl turbulent plasma, the formation of the power-law spectrum shape takes very long time
and may never happen because of the nonlinearity. We propose another ansatz to describe the spectrum shape at finite time.

\end{abstract}


\maketitle

\section{Introduction}

 Magnetic fields are observed in a great variety of astrophysical objects of different scales,
 including stars and interstellar medium, galaxies and galaxy clusters. The origin of most of these
 fields is related to turbulent dynamo mechanism; the conventional point of view is that the
 amplification of seed small-scale magnetic field fluctuations is caused by their advection
 in  a turbulent flow of the conducting fluid or plasma
  ~\cite[see, e.g.,][]{Zeld_book, Moffat, Parker, Brandenburg_rev, Astro-vse}. 
The statistical stationarity 
can be achieved at late stages of evolution as a result of non-linear interaction between the
magnetic field and the flow;
 to the contrary, the most rapid increase of magnetic field takes place at
the kinematic stage when the Lorentz force and the feedback of magnetic field are negligible
~\citep{Kulsrud92}. During this process, the flow is purely hydrodynamic. It obeys the Kolmogorov
theory, which implies the existence of energy flux from larger to smaller scales ~\citep{Frisch}.
Mathematically, this results in nonzero third-order longitudinal velocity correlator  ~\citep{K41};
physically, this means statistical irreversibility of the turbulent flow at all scales
~\cite[e.g.,][]{BiferaleRev}.

Flows with  high magnetic Prandtl numbers are typical for many astrophysical problems: e.g., in
interstellar and intercluster media $Pr_m$ varies from $10^{10} $ to $10^{22}$ (\citep{Han, Sch-Cartesio,
Rincon, PlunianFrick}).
 Kazantsev-Kraichnan model ~\citep{Kazantsev, KraichnanNagarajan} is the most appropriate and
conventional tool to investigate the magnetic field evolution in such flows. It allowed to
calculate magnetic field correlators ~\citep{Kazantsev, Chertkov} and to analyze the spectrum
evolution ~\citep{Kulsrud92, Kazantsev, KraichnanNagarajan, Sch-Cartesio}. But this model assumes
that velocity field is Gaussian (and, hence, time-symmetric) and delta-correlated in time.  Several
modifications of the model were proposed  to take account of finite correlation time
~\cite[e.g.,][]{tau-Sch, tau-Rog, tau-DNS, tau-Bhat14}. In \cite{BalkFalk} the account of the third order
correlator was performed for the equation with additive noise (e.g., the driving force was
considered non-Gaussian).

In this paper we develop the generalization of the Kazantsev-Kraichnan model for time asymmetric
velocity statistics. The $V^3$ model was first proposed in \cite{Scripta19,JOSS1}; it describes
small time anisotropy by taking account of the non-zero third order correlator. Here we combine it
with the Kazantsev approach and calculate the magnetic energy spectrum produced by a slightly
irreversible flow (which corresponds to real hydrodynamic turbulence).

Our consideration is restricted to high magnetic Prandtl numbers, and we consider a wide range of
wave numbers
$$k_\nu \ll k \ll k_d \sim Pr_m^{1/2} k_\nu  $$
where $k_\nu$ and $k_d$ are viscous dissipation and magnetic diffusion characteristic wave numbers,
respectively ~\cite[see][]{Brandenburg_rev, Rincon}. We show that the existence of energy cascade results in more gradual
slope of the spectrum: $\propto k^{1.1}$,  as compared to $k^{3/2}$ predicted by the 'classical'
Kazantsev model. This agrees with the results of numerical simulations \cite{Verma}. However,
estimates show that for  high magnetic Prandtl numbers demanded in astrophysical problems, the
characteristic time needed to saturate the power spectrum  up to the largest (diffusive) wave
number $\sim k_d$  is very long and practically unattainable. So, the power spectrum is only valid
at either infinite time or rather small  (but still $\gtrsim k_{\nu}$) wavenumbers. We propose a
more complicated ansatz to fit the spectrum profile.

\section{Equation for magnetic spectrum}


 We start from the classical problem statement.
 Kinematic transport of  magnetic flux density $B(t,{\bf r})$ advected by random statistically
homogenous and isotropic non-divergent velocity field ${\bf v}({\bf r},t)$, $\nabla \cdot {\bf v}
=0$, is described by the evolution equation
\begin{equation}\label{evol}
\frac{\dd}{\dd
t}\mathbf{B}(\mathbf{r},t)+(\mathbf{v}\nabla)\mathbf{B}-(\mathbf{B}\nabla)\mathbf{v}=
\kappa\Delta\mathbf{B}.
\end{equation}
where $\kappa$ is the diffusivity. The random process ${\bf v}({\bf r},t)$ is assumed to be
stationary, and to have given statistical properties. The initial conditions for magnetic field are
also stochastically isotropic and homogenous. The aim is to find statistical characteristics of the
process $\bf B$, in particular, its pair correlation function.

To get the equation for the pair correlator, one has to take the tensor product of (\ref{evol}) and
$\mathbf{B}(\mathbf{r}')$, and take the average over different realizations of the velocity field.
The result contains cross-correlations of magnetic field and velocity. In the Kazantsev model the
velocity field is assumed to be Gaussian and $\delta$-correlated; so, these cross-correlations are
split by means of the Furutsu-Novikov theorem \citep{Furutsu,Novikov}.
To account
higher-order correlations of velocity, one can use the generalization of this theorem for arbitrary
statistics of $\bf v$ \citep{Novikov,Klyatskin}:
\begin{multline}\label{split_gen}
\langle v_p(\mathbf{r},t) g[\mathbf{v}]\rangle=\sum_{n=0}^{\infty}\frac{1}{n!}\int K_{p\, i_1 \dots
i_n}( \mathbf{r}, t, \mathbf{r}_1, t_1,\dots , \mathbf{r}_n, t_n)
\\
\times\left\langle\frac{\de^n g[\mathbf{v}]}{\de v_{i_1}(\mathbf{r}_1,t_1)\dots \de
v_{i_n}(\mathbf{r}_n,t_n)}\right\rangle\mathrm{d}\mathbf{r}_1\mathrm{d}t_1\dots \mathrm{d}
\mathbf{r}_n\mathrm{d}t_n ,
\end{multline}
where $g[{\bf v}]$ is some functional of ${\bf v}({\bf r},t)$ (in our case it contains second order
combinations of  $\bf B$), $\frac{\delta}{\delta v_i}$ is the functional derivative, and $K_{p\,
i_1 \dots i_n}( \mathbf{R}, t, \mathbf{r}_1, t_1,\dots , \mathbf{r}_n, t_n)$ are $n+1$-th order
connected correlators (cumulants) of velocity:
$$K_{p\, i_1 \dots i_n}( \mathbf{r}, t,
\mathbf{r}_1, t_1,\dots , \mathbf{r}_n, t_n)=\tfrac{\de^{n+1}
\ln\left\langle\mathrm{e}^{\mathrm{i}\int\mathrm{d}\mathbf{r}'\,\mathrm{d}t\,\mathbf{v}(\mathbf{r}',
t)\,\mathbf{y}(\mathbf{r}', t)}\right\rangle}{\de y_{p}(\mathbf{r}, t)\de y_{i_1}(\mathbf{r}_1,
t_1)\dots \de y_{i_n}(\mathbf{r}_n, t_n)}\biggr|_{\mathbf{y}=0}
$$
We are interested in the influence of the third order correlator. So, in the frame of the $V^3$
model \citep{JOSS1, Scripta19, ApJ21} we consider the non-zero second and third order correlators,
\begin{align*}
&K_p( \mathbf{r}, t)=0,
\\
&K_{p\,i_1}( \mathbf{r}, t, \mathbf{r}_1, t_1)=\langle v_p(\mathbf{r}, t)\,
v_{i_1}(\mathbf{r}_1,t_1)\rangle,
\\
&K_{p\,i_1\,i_2}( \mathbf{r}, t, \mathbf{r}_1, t_1, \mathbf{r}_2, t_2)=\langle v_p(\mathbf{r}, t)\,
v_{i_1}(\mathbf{r}_1,t_1)\, v_{i_2}(\mathbf{r}_2,t_2)\rangle
\end{align*}
and we neglect the contribution of the higher-order correlators.
 A vice of this simplification
 is that the probability density is negative in some range
 of its argument
 as only a finite number of  connected correlators are unequal to zero \citep{
 MY}.
This artefact can be fixed in the case of small $F$ by addition negligibly small but non-zero
higher-order correlators. These higher-order corrections would not affect the magnetic field
increment and the slope of the spectrum.

Furthermore, we replace an arbitrary finite-correlation time process by the corresponding
$\delta$-process.  The reason for such substitution is that in the equation with multiplicative
noise, the higher order connected correlators of the noise contribute to the long-time statistical
properties of the solutions only via their integrals (see, e.g., Appendix A in \cite{ApJ21}). So,
the $V^3$ model considers both second and third order correlators as $\delta$-correlated in time:
\begin{align}
&\langle v_i(\mathbf{R}, t)
v_j(\mathbf{R}-\mathbf{r},t-\tau)\rangle=D_{ij}(\mathbf{r})\,\delta_\eps(\tau),
\\
\notag
&\langle v_i(\mathbf{R}, t) v_j(\mathbf{R}-\mathbf{r}_1,t-\tau_1)
v_k(\mathbf{R}-\mathbf{r}_2,t-\tau_2)\rangle
\\
\label{triple}
&\,\,\,\,\,\,\,\,\,\,\,\,\,\,\,\,\,\,\,\,\,\,\,\,\,\,\,\,\,\,\,\,\,\,\,\,\,\,\,\,\,\,\,\,\,\,\,\,\,\,\,\,\,\,\,\,\,\,\,\,\,\,\,\,\,=F_{ijk}(\mathbf{r}_1,\mathbf{r_2})
\phi(\tau,\tau_1,\tau_2)
\\
&\phi=\frac 13 \bigl(
\delta_\eps(\tau_1)\delta_{\eps}(\tau_2)+\delta_\eps(\tau_1)\delta_{\eps}(\tau_2-\tau_1)
\notag
\\
&\,\,\,\,\,\,\,\,\,\,\,\,\,\,\,\,\,\,\,\,\,\,\,\,\,\,\,\,\,\,\,\,\,\,\,\,\,\,\,\,\,\,\,\,\,\,\,\,\,\,\,\,\,\,\,\,\,\,\,\,\,\,\,\,\,+\delta_\eps(\tau_2)\delta_{\eps}(\tau_2-\tau_1) \bigr)
\notag
\end{align}
Here we also take account of statistical homogeneity of $\bf v$. The function  $\delta_{\eps}(t)$
is the regularized $\delta$-functions, i.e., time-symmetric function with narrow support $\eps$
that satisfy $\int \delta_{\eps}(t)dt=1$. This regularization is needed for correct multiplication
of these functions by the $\delta$-functions appearing from variational derivatives; after
calculation of the convolution, we set $\eps=0$. The details of the procedure are described in
 \cite{ApJ21}.  The complicated form of $\phi(t)$  in (\ref{triple}) preserves the
permutation symmetry of the multipliers within the average brackets.

Taking the time integrals in (\ref{split_gen}),  we arrive at
\begin{align}\notag
&\langle v_p(\mathbf{r},t) g[\mathbf{v}]\rangle=\frac{1}{2}\int
D_{ij}(\mathbf{r}-\mathbf{r}')\left\langle\tfrac{\de g[\mathbf{v}]}{\de
v_{j}(\mathbf{r}',t)}\right\rangle\mathrm{d}\mathbf{r}'+
\\\label{split_v3}
+&\frac{1}{6}\int F_{ijk}(\mathbf{r}-\mathbf{r}',\mathbf{r}-\mathbf{r}'')\left\langle\tfrac{\de^2
g[\mathbf{v}]}{\de v_{j}(\mathbf{r}',t)\de
v_{k}(\mathbf{r}'',t)}\right\rangle\mathrm{d}\mathbf{r}'\mathrm{d}\mathbf{r}''
\end{align}

The tensors $D_{ij}$ and $F_{ijk}$ are not arbitrary. They are restricted by the requirements of
isotropy and non-divergency of the flow; also, since we consider the viscous range of scales ($k\gg
k_{\nu}$), the velocity can be treated as a linear function of distance. These conditions reduce
the freedom of each of the two tensors to one constant multiplier. In concordance with
\cite{ApJ21}, we choose the normalization by fixation of two constants:
\begin{align}
&D=-\frac{3}{4}\frac{d^2}{d r^2}\left(\frac{1}{r^2}r_i r_j D_{ij}\right)\Bigr|_{\mathbf{r}=0},
\\
&F=-\frac{3}{4}\frac{\dd^3}{\dd r_1 \dd r_1' \dd r_1''}F_{111}(\mathbf{0},\mathbf{0})
\end{align}
The time scale $D^{-1}$ is of the order of the eddy turnover time at the viscous scale; $f=F/D$
reflects the time asymmetry of the flow.

The applicability of the $V^3$ model formally requires $F\ll D$ (in order to provide positive
probability density by means of negligibly small higher-order contributions). On the other hand, in
\cite{JOSS1} it was shown that the constants $D$ and $F$ are related to the Lyapunov indices
\citep{Oseledets} of the flow, %
 namely,
$$
\frac{\lambda_2}{\lambda_1} = \frac{2F}{2D-F}
$$
(A relation between the second Lyapunov index and the third-order velocity correlator was also
considered in \cite{BF}.)
 The numerical simulation of isotropic turbulence ~\cite{GirimajiPope} and the experiment
~\cite{Luthi} give the ratio of Lyapunov exponents in hydrodynamic flow $\lambda_2/\lambda_1 \simeq
0.14$. This leads to
\begin{equation} \label{0-13}
f \equiv F/D \simeq 0.13
\end{equation}
So, we see that in real turbulence the ratio $F/D$ is small enough to use the $V^3$ model but may
be  essential for  magnetic correlators evolution.

Returning to the equation (\ref{evol}), we note that the magnetic field is also statistically
homogenous, isotropic and non-divergent. This reduces its second order correlator to one scalar
function:
\begin{align}
&\langle B_i(\mathbf{R}, t) B_j(\mathbf{R}+\mathbf{r},t)\rangle\notag
\\\label{BiBj}
&=G(r,t)\de_{ij}+\tfrac{1}{2}r
G'(r,t)\left(\de_{ij}-n_i n_j\right)
\end{align}
Now, multiplying (\ref{evol}) by ${\bf B}({\bf r}',t)$ and taking average, making use of
(\ref{split_v3}), we eventually get
\begin{align}\label{G_evol}
\frac{\dd G(r,t)}{\dd
t}&=\tfrac{2}{3}D\left(r^2G_{rr}''+6rG_r'+10G\right)
\\
&+\tfrac{1}{9}F\left(2r^3G_{rrr}'''+21r^2G_{rr}''+
14rG_r'-70G\right)
\notag
\\
&+2\kappa\left(G_{rr}''+\tfrac{4}{r}G_r'\right)
\notag
\end{align}
The detailed derivation of this equation and the analysis of its grown modes is performed in
\cite{ApJ21}. To consider the  magnetic energy spectrum, we proceed to the Fourier transform of
${\bf B}({\bf r},t)$,
$$
\mathbf{B}(\mathbf{R}, t) =\int\widetilde{\mathbf{B}}(\mathbf{k}, t)
\mathrm{e}^{\mathrm{i}
  \mathbf{k}\mathbf{R}}\mathrm{d}\mathbf{k}
$$
and the magnetic correlation function
\begin{align}\notag
&\langle \widetilde B_i(\mathbf{k}, t) \widetilde B_j(\mathbf{k}',t)\rangle
\\\label{BiBj-2}
&=\frac{M(k,t)}{4\pi
k^2}\left(\de_{ij}-\frac{k_i k_j}{k^2}\right)\de(\mathbf{k}+\mathbf{k}'),
\end{align}
where the function $M$ is related to $G({\bf r},t)$ by
\begin{equation}    \label{G(M)}
G(r,t)=2\int_0^{\infty}\left(-\frac{\cos kr}{(kr)^2}+\frac{\sin
kr}{(kr)^3}\right)M(k,t)\mathrm{d}k.
\end{equation}
 Then the equation (\ref{G_evol}) transforms to
 \begin{align}\label{M_evol}
\frac{\dd M(k,t)}{\dd t}&=\tfrac{2}{3}D\left(k^2M_{kk}''-2kM_{k}'+6M\right)
\\
&+
\tfrac{1}{9}F\left(-2k^3M_{kkk}'''+3k^2M_{kk}''+34kM_{k}'-54M\right)
\notag
\\
&-2\kappa k^2M
\notag
\end{align}
The case $F=0$ corresponds to the Kazantsev-Kraichnan model and was derived and solved in
\cite{Kazantsev, Kulsrud92}. Here we investigate the corrections produced by the term
responsible for time asymmetry. We consider the long-time evolution of the Green function:
\begin{equation}
M(k,0)=\de(k-k_0).
\end{equation}
By the change of variables
$$
 \xi = \ln k
$$
the Eq.(\ref{M_evol}) can be reduced to
 \begin{align}
\frac{\dd M(\xi,t)}{\dd
t}&=\tfrac{2}{3}D\left(M_{\xi\xi}''-3M_{\xi}'+6M\right)
\notag
\\
&+
\tfrac{1}{9}F\left(-2M_{\xi\xi\xi}'''
+9M_{\xi\xi}''+27M_{\xi}'-54M\right)
\notag
\\\label{M-evol-xi}
& -2\kappa e^{2\xi} M
\\
M(\xi,0)&=\mathrm{e}^{-\xi_0}\,\de(\xi-\xi_0) \notag
\end{align}

\begin{figure}[b]
\center{\includegraphics[width=.97\linewidth]{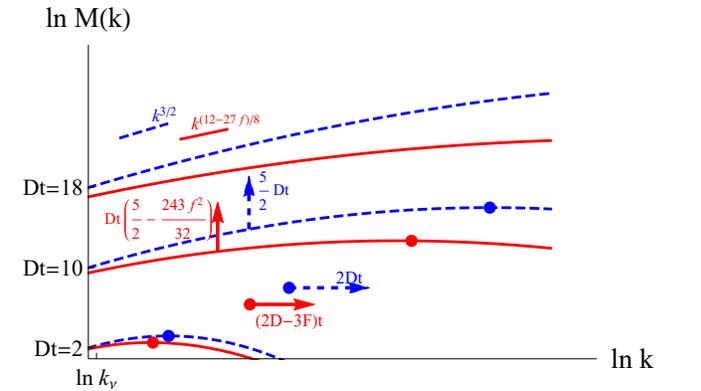}} \caption{\label{SpectrumIdeal}
The evolution of magnetic energy spectrum in the non-diffusivity approximation: the
Kazantsev-Kraichnan model (blue dashed lines) and the $V^3$ model for $f=0.13$ (\ref{0-13}) (red
solid lines). The thick points correspond to the maxima of the spectra.  The time moments
(normalized by $D^{-1}$) are indicated along the vertical axis.
 }
\end{figure}
\section{Zero diffusivity }

First, consider the limit of zero magnetic diffusivity: $\kappa=0$. By means of the Fourier
transform, we find the formal solution:
\begin{align} 
M(&\xi,t)=\frac{\mathrm{e}^{-\xi_0}}{2\pi}
\int\mathrm{d}\eta\,\mathrm{e}^{-\mathrm{i}\,\eta\,(\xi-\xi_0)}
\notag
\\
&\times\mathrm{e}^{t\,\left[\frac{2}{3}D(-\eta^2+3\mathrm{i}\,\eta+6)+\frac{1}{9}F(-2\mathrm{i}\,\eta^3-9\eta^2-27\mathrm{i}\,\eta-54)\right]}
\label{M(xi)}
\end{align}
The integral diverges formally at large $\eta$; this is an artefact of the $V^3$ model, this
divergence is the result of the 'parasite' solution that is produced by the third-order term and
tends to infinity as $F \to 0$. The account of higher order terms would evidently eliminate this
divergence.

To calculate the integral, we use the saddle-point method. The equation for the saddle point is
$$
-\tfrac{\mathrm{i}\,\de\xi}{t}+D\left(-\tfrac{4}{3}\eta^*+2\,\mathrm{i}\right)+F\left(-\tfrac{2}{3}\mathrm{i}\,{\eta^*}^2-2\,\eta^*-3\,\mathrm{i}\right)=0
$$
where
$$
\de \xi = \xi - \xi_0
$$
The 'physical' solution is the one that is close to the Kraichnan solution corresponding to $F=0$:
\begin{equation}
\eta^*=\mathrm{i}\left(\frac{3}{2}+\frac{1}{f}\left(1-\sqrt{1+\frac{27}{4}f^2+f\,\frac{3\xi}{2Dt}}\right)
\right).
\end{equation}
Substituting this into (\ref{M(xi)}) we obtain:
\begin{widetext}
\begin{equation}\label{M(xi)_id}
M(\de\xi,t)\propto\mathrm{exp}\left[\frac{3\de\xi}{2}+\frac{1}{18f^2}\left(2(4+63f^2)Dt+18f\de\xi
-\frac{\left((4+27f^2)Dt+6f\de\xi\right)^{3/2}}{\sqrt{Dt}}\right)\right]
\end{equation}
Returning from $\xi$ to $k$ and leaving only the second order in $f=F/D$ (which restricts us to
$\xi/(Dt) \le f^{-2}$)  we finally get
\begin{align}
\label{M(k)_id} &M(k,t)\simeq\frac{1+{f^2}/{8t}} {2\sqrt{\pi t} {k_0}}
\mathrm{e}^{\left(\frac{5}{2}-\frac{243}{32}f^2\right)Dt}
\left(\frac{k}{k_0}\right)
^{\frac{3}{2}-\frac{27}{8}f}\,\mu_{id}\left(\frac{k}{k_0},t\right),\,\,\,\,k<k_0\mathrm{e}
^{\frac{D}{f^2}t}
\\
\nonumber
\mu_{id}(y&,t)=\mathrm{exp}\left[-\left(1-\frac{27}{8}f^2\right)\frac{3\ln^2y}{8Dt}+\frac{3f}{32}\frac{\ln^3y}{(Dt)^2}-\frac{27f^2}{512}\frac{\ln^4y}{(Dt)^3}\right].
\end{align}
\end{widetext}

This solution describes the evolution of spectrum in the absence of magnetic diffusivity. Its
behavior and its comparison with the result of the  Kazantsev-Kraichnan model
\citep{Kulsrud92,Sch-Cartesio} is illustrated in Fig.~\ref{SpectrumIdeal} where we choose
$k_0=k_{\nu}$.
 We see
that the spectrum grows exponentially (the first multiplier), and  it exponentially spreads into
the region of large wave numbers (the multiplier $\mu_{id}$). This situation coincides
qualitatively with the one obtained in the Gaussian case. However, the slope of the spectrum is
more flat than that in the Kraichnan model, the exponential growth is slower,  and the speed of the
maximum of the spectrum is smaller (for $F>0$). Thus, the magnetic energy increases significantly
slower than in the Gaussian case. Indeed, taking the integral of (\ref{M(xi)_id}) we have:
$$
E(t)=\int M(k,t)\mathrm{d}k\propto\mathrm{e}^{t\,(\frac{20}{3}D-\frac{70}{9}F)}  
$$
\begin{figure*}
\begin{minipage}[h]{0.47\linewidth}
\center{\includegraphics[width=1\linewidth]{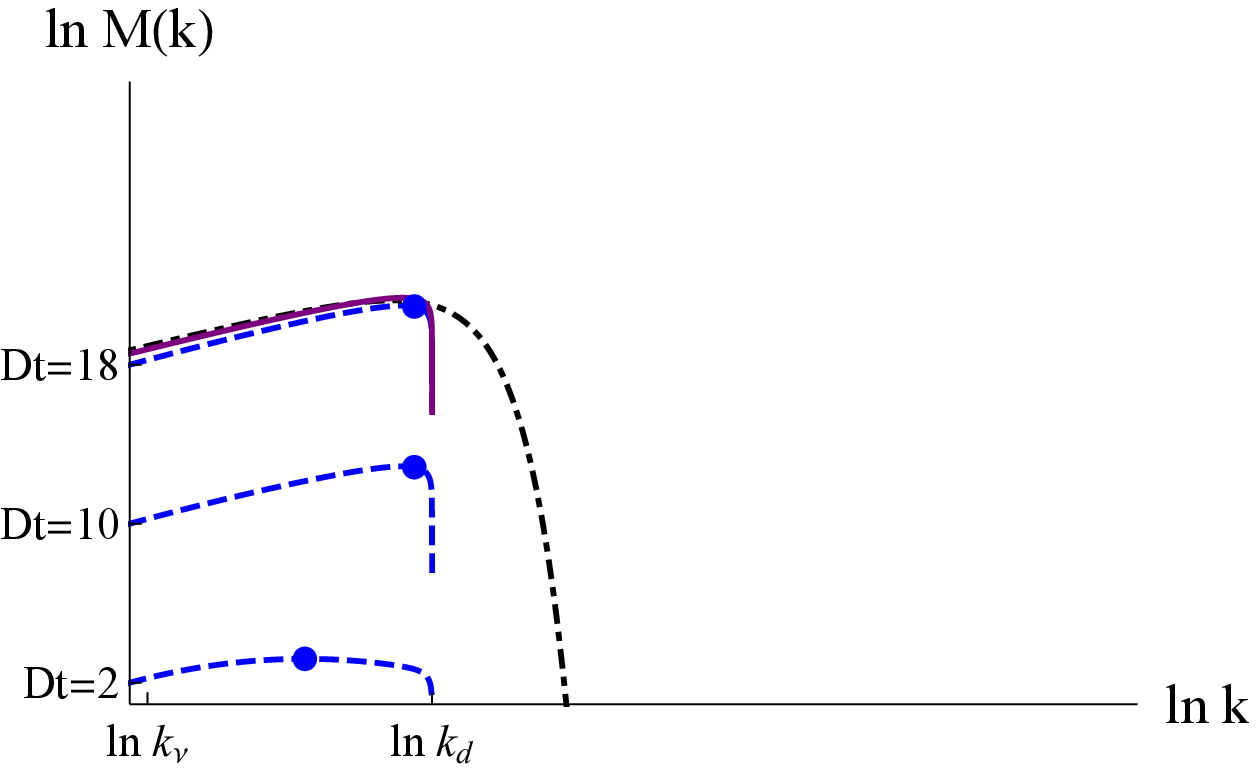}} a) \\
\end{minipage}
\hfill
\begin{minipage}[h]{0.47\linewidth}
\center{\includegraphics[width=1\linewidth]{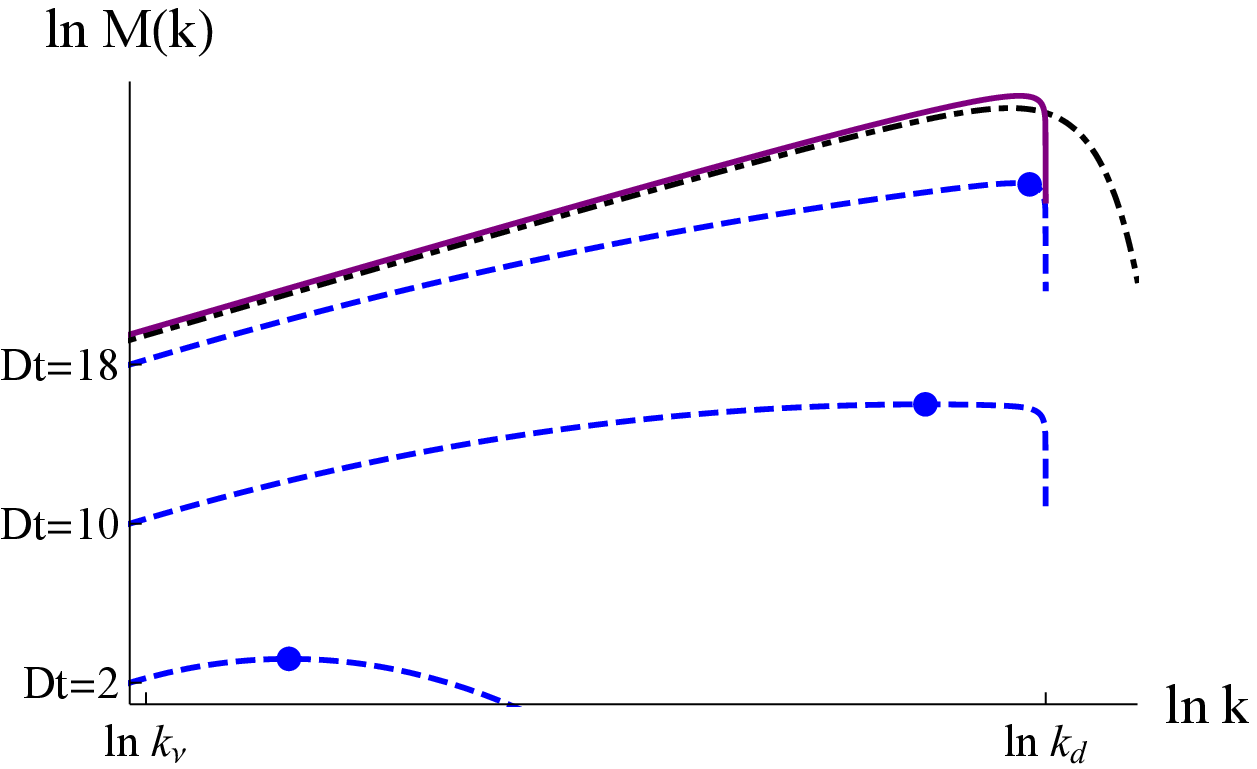}} \\b)
\end{minipage}
\vfill
\begin{minipage}[h]{0.47\linewidth}
\center{\includegraphics[width=1\linewidth]{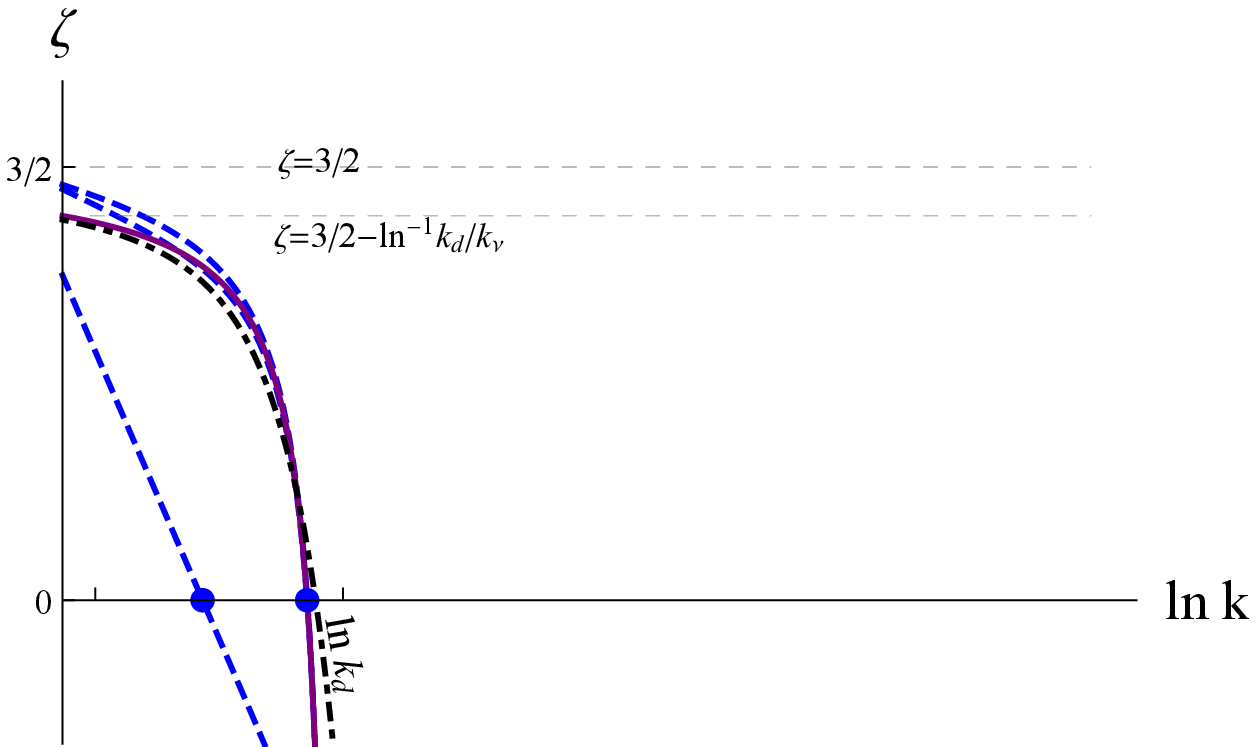}} c) \\
\end{minipage}
\hfill
\begin{minipage}[h]{0.47\linewidth}
\center{\includegraphics[width=1\linewidth]{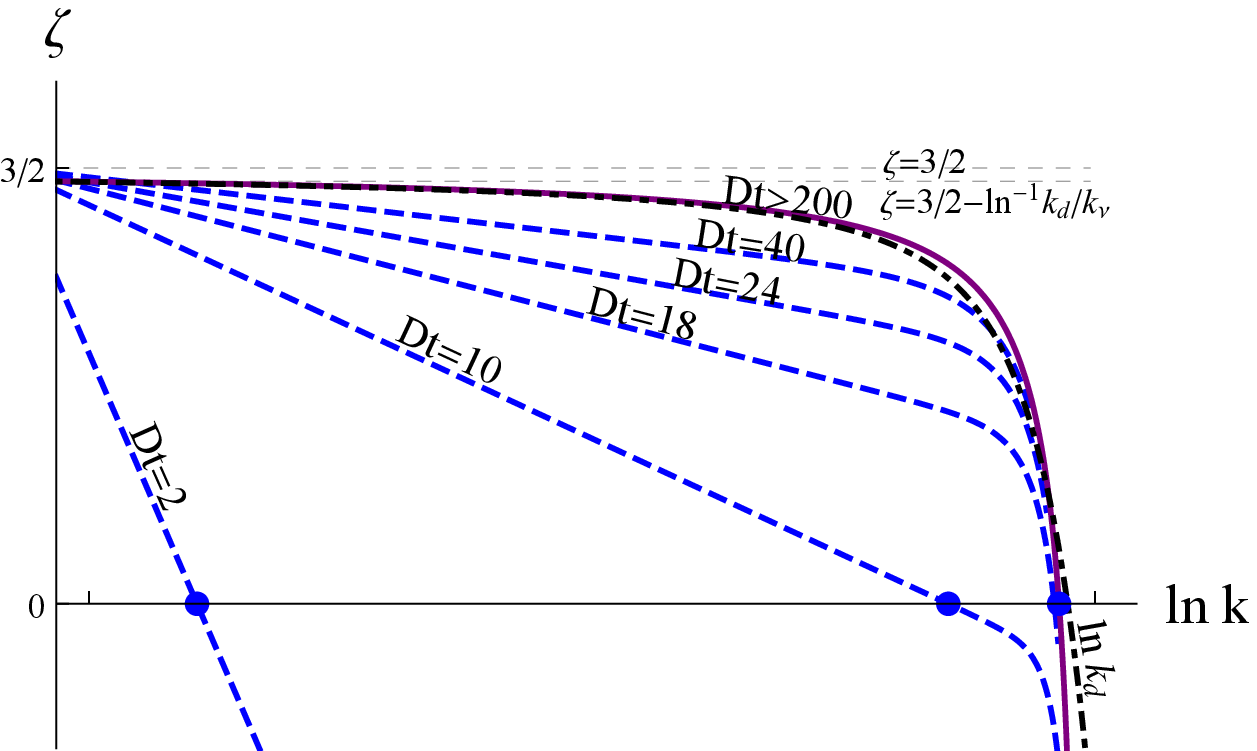}} d) \\
\end{minipage}
\caption{\label{Scaling} Evolution of the spectrum $M$ and the logarithmic derivative $\zeta$
(\ref{dlnMdlnk})
 in the Kazantsev-Kraichnan model with diffusion
taken into account by means of zero boundary condition at $k=k_d$ ('absorption' model),
(\ref{M(k)_d}): (a),(c) $Pr_m=10^3$; (b),(d)  $Pr_m=10^{10}$. The scale is the same for all
figures. The  solid purple line corresponds to the stationary shape of the spectrum in the
'absorption' model ~(\ref{M(k)_st}), the dashed-dotted black line represents the exact solution for
the stationary shape of the spectrum  ~\citep{Kazantsev}; both of them are normalized by $Dt=18$.
The thin dashed horizontal lines in the figures (c) and (d) correspond to the theoretical
predictions for the power of the self-similarly growing spectrum for the Kazantsev model with
$\kappa =0$ and for the 'absorption' model (\ref{M(k)_st}) at $k=k_{\nu}$. }
\end{figure*}

\section{Spectrum formation}

However, for much higher Prandtl numbers it appears that the stabilization time needed to reach the
self-similar growth is extremely long; the non-linear stage would start much earlier than the
stationary shape of the  spectrum is formed. To prove this, we now restrict ourself (for brevity)
by the Kazantsev-Kraichnan  approximation.

The diffusive term in the equation (\ref{M-evol-xi}) suppresses the magnetic energy at $k \gtrsim
k_d$ and does not effect much the smaller wave numbers. In the variables $\xi$ this boundary is
sharp, and, to take diffusivity into account, it is natural to simplify the problem by full
suppression of the magnetic field at the wave numbers $\ge \xi_d$ and by neglecting diffusivity at
smaller $\xi$. So, we consider the  'absorption' model  with the boundary condition $M=0$ for $\xi
\ge \xi_d$ and with $\kappa=0$ for $\xi<\xi_d$.

Calculating the Green function of the 'non-diffusive' equation (\ref{M-evol-xi}) for this
'absorption' boundary condition and returning to the $k$ variables, we obtain (for $f=0$):
\begin{align}\label{M(k)_d}
M(k,t)=\frac{1}{2\sqrt{\pi\,t}}&\frac{\mathrm{e}^{\frac{5}{2}Dt}}{k_0}
\left(\frac{k}{k_0}\right)^{\frac{3}{2}}\,\left(\mathrm{exp}\left[-\frac{3}{8Dt}\ln^2\frac{k}{k_0}\right]\right.
\notag
\\
-&\left.\mathrm{exp}\left[-\frac{3}{8Dt}\ln^2\frac{k\,k_0}{k_d^2}\right]\right),\,\,\,
k<k_d
\end{align}
 The phase of stationary  self-similar growth of the  spectrum means that the term in
the brackets becomes roughly independent of $k$, i.e., the exponents are smaller than unity. Then
the term in the brackets can be expanded into a Taylor series to the first order for all
$k_\nu<k<k_d$:
\begin{equation}\label{M(k)_st}
M(k,t)\simeq\frac{3}{4\sqrt{\pi
}D\,t^{3/2}}\frac{\mathrm{e}^{\frac{5}{2}Dt}}{k_0}\left(\frac{k}{k_0}\right)^{\frac{3}{2}}
\ln\frac{k_d}{k_0}\ln\frac{k_d}{k},\,\,\,\,\,\left\{\mycom{
k<k_d,}{ \,t \gtrsim t_{st}}\right\}.
\end{equation}
Here $t_{st}$ is the time when self-similarity of the spectrum is established; it is determined by
the exponents in the brackets,
 $$
t_{st}= \frac{3}{8D}\ln^2 \frac{k_d}{k_0} \sim \frac{3}{8D}\ln^2 Pr_m
 $$
 But this time appears to be very long for large magnetic Prandtl numbers: for instance,
for $Pr_m \sim 10^{10}$  (which is moderate value for cosmic plasma) the characteristic time of the
spectrum shape stabilization is $Dt_{st}\simeq 200$. The energy would increase  by 200 orders of
magnitude during this time! It is evident that the nonlinear feedback of magnetic field would start
earlier than the spectrum saturates.

In Fig.~\ref{Scaling},a,b we show the evolution of the spectrum (\ref{M(k)_d})  for two different
Prandtl numbers: $Pr_m=10^3$ (a) and $Pr_m=10^{10}$ (b). For definiteness, we take $k_0=k_{\nu}$.
We compare the solutions with the exact stationary growing solution  of the Kazantsev equation. One
can see that the self-similarly growing solution found from the 'absorption' model practically
coincides in the region $k<k_d$ with the exact solution, which proves the validity of the
'absorption' approximation. Furthermore, in accordance with our expectations, the
'moderate-Prandtl' spectrum corresponding to $Pr_m=10^3$ is practically saturated at $t=18D^{-1}$,
while the high-Prandtl spectrum with $Pr_m=10^{10}$ is far from the saturation after the same time.
To concentrate upon the slope of the spectrum, we plot dependence of  the logarithmic  derivative
on the wave number,
\begin{equation}  \label{dlnMdlnk}
 \zeta(k)=\frac{\mathrm{d}\ln M(k)}{\mathrm{d} \ln{k}}
\end{equation}
for both values of the Prandtl number (Fig.~\ref{Scaling} c,d). One can see that the power law
is rather rough approximation. Indeed, in the case of $Pr_m=10^3$ the spectrum saturates quickly
but its shape is far from a power law. In particular, only in the longest-wave region $k \simeq
k_{\nu}$ one can find the power predicted from the stationary-growing solution. For $Pr_m=10^{10}$,
the scaling region is wide but the power of the spectrum does not coincide with that predicted from
the Kazantsev equation for any reasonable time. 

The approximation of zero diffusivity is only valid until the maximum of the spectrum reaches the
diffusion scale $k_d$:
$$
t \le t_d \sim D^{-1} \ln (k_d/k_0)
$$
At longer time, the expansion of the spectrum into the short-wave region  stops; however, the
spectrum continues to grow exponentially inside the region $k<k_d$. The conventional understanding
is that, after some time, the power law shape would propagate from the long-wave to the short-wave
part of the spectrum, and the spectrum would become self-similar and stationary-growing. This
self-similarity would last until the  dynamo becomes non-linear and starts acting back on the
velocity field.  In ~\cite{Sch-DNS} this picture is proved numerically for Prandtl numbers
$\lesssim 10^3$.

\begin{figure}
\center{\includegraphics[width=.97\linewidth]{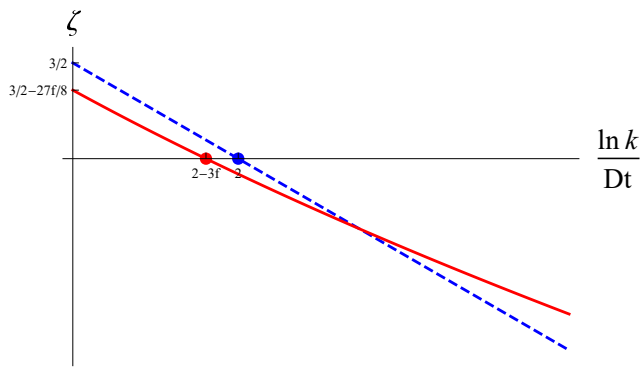}}
 \caption{\label{ScalingUniv} Dependence of the logarithmic derivative (\ref{zeta(k)_anz}) on its
 only argument
 $\ln (k/k_0)/Dt$: Kazantsev-Kraichnan model (dashed blue  line) and the $V^3$ model with $f=0.13$
 (solid red line). The points indicate the maxima of the spectra. }
\end{figure}

While time is smaller than the saturation time $t_{st}$, the  second term in the brackets in
(\ref{M(k)_d}) remains much smaller than the first term; hence, this during this time, the effect
of the magnetic diffusivity at relatively large values of $k$ ($k \sim k_0$) remains negligible.
So, the solution (\ref{M(xi)_id})  found in the approximation of zero diffusivity for $k_\nu\gtrsim
k_0\gg k_d$ appears to be valid much longer than $t_d$. The logarithmic derivative of
(\ref{M(xi)_id}) is
\begin{align}
\label{zeta(k)_anz}
\zeta=\left[\begin{aligned}&\frac{3}{2}-\frac{3}{4Dt}\ln \frac{k}{k_0} &&,\,f=0
\\
&\frac{3}{2}-\frac{1}{f}\left(\sqrt{1+\frac{27}{4}f^2+\frac{3}{2}f\frac{\ln
k/k_0}{Dt}}-1\right)&&,\,f\ne0\end{aligned}\right.
\end{align}
One can see that   $\zeta$  depends on time and wave number only in the combination $t^{-1}\ln
(k/k_0)$. The dependence has a universal form (Fig.~\ref{ScalingUniv}) and is completely determined
by the asymmetry parameter $f$. This property can be used to fit the experimental and numerical
spectra in a more accurate way than the power fit.

\section{Discussion}

So, in this paper we consider the evolution of magnetic energy spectrum produced by turbulent
dynamo in the viscous range of scales, in the case of high  Prandtl numbers, $Pr_m\gg 1$. We  find
a non-trivial correction to the Kazantsev spectrum $k^{3/2}$ produced by the non-zero third-order
velocity correlator. We also propose the finite-time ansatz (\ref{zeta(k)_anz}) for the spectrum
shape that may be more accurate and may better correspond the numerical  results than the power-law
spectrum.

Why is the third order correlator essential? The Kazantsev's result is very stable: the effect of
finite correlation time was considered in ~\cite{tau-DNS} and it was shown that the account of
finite time correlation does not change the slope of the spectrum. In \cite{tau-Bhat14} the
time-symmetric non-Gaussianity with small correlation time
was also shown to be insignificant. Here we see that, to the
contrary, the non-zero third order velocity correlator brings a significant change to the spectrum
power. Probably, this peculiar effect of the third-order correlator is provided by its relation to
the time asymmetry of the flow. Indeed, in time symmetric flows the correlations of odd orders do
necessarily equal zero
(because of the invariance with respect to the inversion of signs of all velocities).  So, the
non-zero third order correlator ensures the existence of the energy cascade. It is natural that it
may play essential role in the advection process.

We note that in two-dimensional incompressible flows, statistical asymmetry of the velocity field
is impossible in the viscous regime  ($\lambda^{2D}_1=-\lambda^{2D}_2$);  for this reason, we
suppose that 
the two-dimensional spectrum calculated in the frame of Kazantsev-Kraichnan model
~\citep{Sch-Cartesio} is also valid for arbitrary statistics, unlike the 3-dimensional case.

The correction to the exponent of the spectrum depends on the asymmetry parameter $f$, $\zeta =
3/2-27/8f$. For numerical estimate, we use the numerical and experimental measurements
~\citep{GirimajiPope, Luthi} of the Lyapunov exponents in Kolmogorov turbulence;
they correspond to the Reynolds numbers $Re\sim 400$ ($Re_{\lambda}\simeq 38$) and  give $f=0.13$
(\ref{0-13}). The resulting spectrum slope then appears to be $k^{1.1}$. For any specific flow, the
exponent would depend on the ratio of the Lyapunov exponents (and thus, indirectly, on the Reynolds
number).  We compare this result with the data of numerical high-Prandtl dynamo simulation in a
high-Reynolds flow  ($Re=9.68\times 10^6$, $Pm=10^3$) performed in the frame of the Shell model
~\citep{Verma}.
In the viscous range, the power law found from the $V^3$ model fits well the numerical data.

 Regrettably, many high-magnetic Prandtl
dynamo simulations are performed with Reynolds numbers of the order of unity ~\cite[e.g.,][]{tau-DNS,
Sch-DNS}; thus, the relation of the Lyapunov exponents in these investigations may differ
essentially from the value (\ref{0-13}) calculated for Kolmogorov turbulence. This makes it
difficult to compare them to our result.
On the other hand, the Shell model simulations~\cite[e.g.,][]{PlunianFrick, Verma} allow to get high magnetic Prandtl numbers together
with high Reynolds numbers, which allows to check the predictions for the dynamo produced in a
turbulent flow with energy cascade.  Numerical simulations in a 'synthetic' velocity field with
prescribed ratio of the Lyapunov indices could be another way to check the predictions of the $V^3$
model. One could investigate both the deviation of the final power spectrum from the '3/2' law of
the Kazantsev's model, and the intermediate behavior of the spectrum before it gets saturated.


\begin{acknowledgments}
The authors are grateful to Professor A.V.~Gurevich for his permanent attention to their work. 
The work of A.V.K. was supported by the RSF Grant No. 20-12-00047.
\end{acknowledgments}

\end{document}